\documentclass[prl,letterpaper,twocolumn,showpacs]{revtex4}
\usepackage{times,xspace}
\usepackage{amsbsy,amssymb,amsmath,bm}
\usepackage{graphicx,color,epsfig}
\usepackage{fancyhdr}

\def\bbbc{{\mathchoice {\setbox0=\hbox{$\displaystyle\rm C$}\hbox{\hbox
to0pt{\kern0.4\wd0\vrule height0.9\ht0\hss}\box0}}
{\setbox0=\hbox{$\textstyle\rm C$}\hbox{\hbox
to0pt{\kern0.4\wd0\vrule height0.9\ht0\hss}\box0}}
{\setbox0=\hbox{$\scriptstyle\rm C$}\hbox{\hbox
to0pt{\kern0.4\wd0\vrule height0.9\ht0\hss}\box0}}
{\setbox0=\hbox{$\scriptscriptstyle\rm C$}\hbox{\hbox
to0pt{\kern0.4\wd0\vrule height0.9\ht0\hss}\box0}}}}

\newcommand{\ignore}[1]{}
\newcommand{\mComment}[1]{}
\newcommand{\gComment}[1]{}
\newcommand{\jComment}[1]{}
\newcommand{\rComment}[1]{}
\newcommand{\lComment}[1]{}
\begin{document}

\title{ Emergent Symmetry and Dimensional Reduction at a Quantum Critical
Point}
\author{J. Schmalian$^{1}$, C. D. Batista$^{2}$}
\affiliation{$^{1}$ Department of Physics and Astronomy, Iowa State University and Ames
Laboratory, Ames IA50011 \\
$^2$ Theoretical Division, Los Alamos National Laboratory, Los Alamos, NM
87545}
\date{\today }

\begin{abstract}
We show that the spatial dimensionality of the quantum critical point
associated with Bose--Einstein condensation at $T=0$ is reduced when the
underlying lattice comprises a set of layers coupled by a frustrating
interaction. For this purpose, we use an heuristic mean field approach that
is complemented and justified by a more rigorous renormalization group
analysis. Due to the presence of an emergent symmetry, i.e. a symmetry of
the ground state that is absent in the underlying Hamiltonian, a
three--dimensional interacting Bose system undergoes a chemical potential
tuned quantum phase transition that is strictly two dimensional. Our
theoretical predictions for the critical temperature as a function of the
chemical potential correspond very well with recent measurements in BaCuSi$
_{2}$O$_{6} $.
\end{abstract}

\maketitle

\section{Introduction}

The universal properties that appear in the proximity of a critical point
are determined by a few relevant properties. The spatial dimensionality, $d$
, is one of them \cite{Stanley}. This is evident from the fact that, in
general, the critical exponents depend on $d$. Correspondingly, for strongly
anisotropic systems of weakly coupled chains or planes, critical behavior
characteristic for $d=1$ or $d=2$ respectively can be observed beyond a
certain distance from the critical point. The critical behavior crosses over
to three--dimensional only in the close vicinity of the critical point of
such anisotropic systems.

In contrast to this conventional dimensional crossover, the spatial
dimensionality can be effectively reduced under certain conditions as the
system approaches the critical point. This phenomenon of \textit{dimensional
reduction} is closely related to the notion of \textquotedblleft \textit{\
emergent sliding symmetries}\textquotedblright\ \cite{Batista04}. Those are
physical systems for which new symmetry transformations appear at low
energies (in some cases only at $T=0$). In other words, the low energy
spectrum of the system Hamiltonian is invariant under these symmetries but
the whole spectrum is not \cite{Batista05}. We call these transformations
\textquotedblleft emergent symmetries\textquotedblright\ because they only
appear at low energies. By \textquotedblleft sliding
symmetry\textquotedblright\ \cite{Zohar05} we mean symmetry transformations
that only change a subset of the degrees of freedom which occupy a region of
dimension lower than $d$. For instance, if our system is a 3D quantum magnet
and it is invariant under a spin rotation restricted to a given layer, such
operation is a \textquotedblleft sliding symmetry\textquotedblright .

A simple example of an emergent sliding symmetry is provided by classical
spins on a body centered tetragonal (BCT) lattice with antiferromagnetic XY
exchange interactions. If the inter--layer exchange interaction, $J_{\perp }$
, is smaller than the intra--layer one, $J_{\parallel }$, the energy is
minimized when the spins are antiferromagnetically aligned on each layer.
Since the staggered magnetization of each layer can point in any arbitrary
direction, the ground state manifold is highly degenerate. In this case, an
arbitrary spin rotation along the z--axis which acts only on the spins of a
given layer is a sliding emergent symmetry. It is a symmetry because it does
not change the ground state energy. It is \textquotedblleft
emergent\textquotedblright\ because it only exists at $T=0$: the energy 
\textit{does not} remain invariant if we apply the same transformation to an
excited state. In particular, this symmetry is the manifestation of a simple
physical property: the order parameters (staggered magnetization) of
different layers are decoupled at zero temperature. Consequently, in spite
of the 3D nature of the system, the antiferromagnetic ordering is 2D at $T=0$
. This is a simple example of dimensional reduction that results from two
key ingredients: the classical nature of the degrees of freedom and the
frustrated nature of the interactions. 
\begin{figure}[tbh]
\vspace*{0.0cm} 
\hspace*{0.3cm} 
\includegraphics[angle=90,width=10cm]{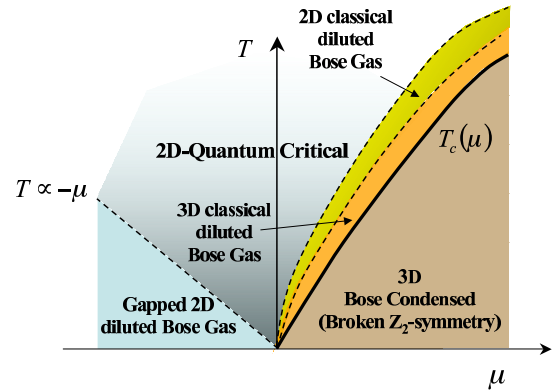}
\vspace{-2.5cm}
\caption{Phase diagram obtained by the renormalization group approach.}
\label{phase_diagram}
\end{figure}

It is natural to ask if the phenomenon of dimensional reduction also exists
in quantum systems. In most of the cases, the emergent sliding symmetry is
removed by zero point fluctuations. For instance, if we consider now the
quantum version of the XY model on a BCT lattice with $J_{\parallel
}>J_{\perp }$, the ground state is no longer invariant under spin rotations
along the z--axis of all the spins on a given layer. Therefore, this
operation is an emergent symmetry only in the classical limit. Zero point
fluctuations remove this symmetry by inducing a finite coupling between the
staggered magnetization on different layers \cite{Maltseva05,Yildirim96}.
This is a particular example of the phenomenon known as \textquotedblleft
order from disorder \cite{Shender82}.

However, zero point fluctuations not always restore the dimensionality by
removing the emergent sliding symmetries. Such symmetries can appear at
special points of the quantum phase diagram and lead to dimensional
reduction. The main characteristic of these special points is that the
ground state becomes ``classical'', i.e., it is a direct product of
eigenstates of a local physical operator. For instance, the fully polarized
ferromagnet is a ``classical'' state for any spin $S$. A simple example of
dimensional reduction in a quantum system is given in Ref.\cite{Batista004}
for a Klein model of S=1/2 spins on square lattice. In that case, the
dimensional reduction from $d=2$ to $d=1$ occurs at a first order quantum
phase transition point. An immediate physical consequence of this
dimensional reduction is the emergence of fractional excitations
characteristic of one--dimensional systems.

We have shown recently that the phenomenon of dimensional reduction can also
occur at a quantum critical point (second order quantum phase transition) 
\cite{Batista07}. For this purpose, we considered the quantum XY magnet of
our previous example but in the presence of an external magnetic field, $H$,
along the Z--direction. The ground state is antiferromgnetic for $H=0$ while
the Zeeman term dominates at high fields leading to a fully polarized ground
state. The antiferromagnetically ordered XY component decreases continuously
as a function of $H$ and vanishes at the critical field $H_{c}$. The spin
system becomes fully polarized for $H>H_{c}$. The corresponding quantum
phase transition is denoted as Bose--Einstein condensation (BEC) because the
order is suppressed by suppressing the \textit{amplitude} of the order
parameter or staggered magnetization. In contrast, the thermodynamic phase
transition is denoted as XY because the order is suppressed by phase
fluctuations. As we will see below, this difference is crucial for the
phenomenon of dimensional reduction. The BEC--QCP of the system under
consideration has a peculiar property: the disordered state for $H>H_{c}$ is
\textquotedblleft classical\textquotedblright\ because it is a direct
product of eigenstates of $S_{i}^{z}$ ($z$--component of the spin operator
on a given site $i$). In other words, the zero point phase fluctuations that
restore the 3D ordering at $H=0$ are no longer present for $H>H_{c}$ simply
because the XY spin component has been suppressed completely. Since the
transition is continuous, the 3D coupling induced by these phase
fluctuations must vanish continuously when $H$ approaches $H_{c}$ from the
ordered side. For this reason, dimensional reduction occurs right at the
critical point.

The specific motivation for the theory presented in this paper is the
unusual temperature dependence of the transition temperature as function of
magnetic field in frustrated magnet BaCuSi$_{2}$O$_{6}$\cite{Suchitra06}. We
describe this system by a Heisenberg Hamiltonian of $S=$ $\frac{1}{2}$ spins
forming dimers on a body--centered tetragonal lattice, closely approximating
the case of BaCuSi$_{2}$O$_{6}$ \cite{Sparta04,Samulon06}. The dominant
Heisenberg interaction, $J\sum_{\mathbf{i}}\mathbf{s}_{i1}\cdot \mathbf{s}
_{i2}$, is between spins on the same dimer $i$. Since there are two low
energy states in an applied magnetic field, the singlet and the $
s_{i1}^{z}+s_{i2}^{z}=1$ triplet, we can describe the low energy sector
either using hard--core bosons, or, in terms of the above mentioned XY
model. In case of the hard core boson description, the triplet state
corresponds to an effective site $i$ occupied by a boson while the singlet
state is mapped into the empty site \cite{Giamarchi99,Jaime04}. The number
of bosons (number of triplets) equals the magnetization along the $z$--axis.
The chemical potential $\mu =g\mu _{B}(H-H_{c1})$ is determined by the
applied magnetic field, $H$, and the critical field $g\mu
_{B}H_{c1}=J-2J^{\prime }$ (where $g$ is the gyromagnetic factor, $\mu _{B}$
is the Bohr magneton and $J^{\prime }$ is the inter--dimer exchange
interaction). The hoppings $t_{\parallel }=J^{\prime }$ and $t_{\perp
}=J^{\perp }$ ($J^{\perp }$ is the frustrated inter--layer exchange
interaction) are determined by the inter--dimer exchange interactions
between spins. Recently, it was shown by R\"{o}sch and Vojta \cite
{Rosch,Rosch2} that the inclusion of the two higher triplet modes generates
a small coherent second neighbor hopping of low energy triplets between
layers $t_{\perp ,2}^{\ast }\simeq J_{\perp }^{6}/J^{5}$. This interesting
effect restores the $d=3$ character of the spin problem due to the fact that
the paramagnetic ground state for $H<H_{c1}$ is not purely classical.
Although it can be described as classical state (direct product of singlets
on each dimer) to a very good approximation, there are small zero--point
phase fluctuations that result from virtual process to the higher triplet
states (creation and annihilation of triplet pairs with zero net magnetic
moment). It was also pointed out in Refs.\cite{Rosch,Rosch2} that the
dimensional reduction is still exact at $H=H_{c2}$ (saturation field)
because the state for $H>H_{c2}$ is purely classical. For realistic values
of $J=49.5(1)$K and $J_{\perp }<J^{\prime }$, $J_{\perp }^{6}/J^{5}<0.1$mK
in BaCuSi$_{2}$O$_{6}$. This implies that the mechanism discussed in our
paper is still dominant for all experimentally accessible temperatures $
T\gtrsim 30$mK. Moreover, the U(1)-symmetry breaking terms induced by
dipolar interactions will produce a crossover to QCP with discrete symmetry
at $T\sim 10$mK \cite{Suchitra006} before the mechanism of Ref.\cite{Rosch}
sets in. Finally, the inevitable presence of finite non-frustrated couplings
in real systems will eventually restore the $d=3$ behavior below some
characteristic temperature $T_{0}<30$mK.

Despite the above mentioned effects, where lattice distortions, dipolar
couplings or excitations to high energy triplets cause a restoration of
three dimensional behavior at very low temperatures, is it important to
stress that the boson model discussed in this paper is a nontrivial
interacting many-body system where the dimensional reduction at the $T=0$
quantum critical point is exact. Materials that can be described in terms of
a chemical potential tuned Bose-Einstein Condensation on a frustrating
lattice are then candidates for the dimensional reduction as caused by an
emergent symmetry in the problem. In this sense are the conclusions of our
paper are not limited to BaCuSi$_{2}$O$_{6}$ alone.

The main purpose of the present work is to derive the critical properties of
the field induced BEC--QCP for the XY magnet mentioned above. The key
finding of our result is the detailed phase diagram of Fig.\ref
{phase_diagram}, where we show the various crossover regimes of a chemical
potential tuned BEC on a frustrated lattice. This work complements the
results presented in Ref.\cite{Batista07} by including a renormalization
group approach (Sec. IV) which provides formal justification for the
heuristic mean field approach presented in Ref.\cite{Batista07} and is
summarized in detail in Section III. The model for the XY-magnet on a BCT
lattice is introduced in Sec. II. For practical reasons, we use the language
of hard core bosons which are equivalent to $S=1/2$ spins after a
Matsubara--Matsuda transformation \cite{Matsubara56}. Our conclusions are
presented in Section IV.

\section{Model}

We start from the Hamiltonian of interacting spinless bosons on a body
centered cubic lattice 
\begin{equation}
H_{B}=\sum_{\mathbf{k}}\left( E_{\mathbf{k}}-\mu \right) a_{\mathbf{\ k}
}^{\dagger }a_{\mathbf{k}}+u\sum_{i}n_{i}n_{i}.  \label{Hb}
\end{equation}
Here $n_{i}=a_{i}^{\dagger }a_{i}$ is the local number operator of the
bosons and 
\begin{equation}
a_{\mathbf{k}}^{\dagger }=\frac{1}{\sqrt{N}}\sum_{i}a_{i}^{\dagger }e^{i 
\mathbf{\ k\cdot R}_{i}},
\end{equation}
the corresponding creation operator in momentum space. The tight binding
dispersion for nearest neighbor boson hopping on the BCT lattice is 
\begin{equation}
E_{\mathbf{k}}=\varepsilon _{\mathbf{k}_{\parallel }}+2t_{\perp }\gamma _{ 
\mathbf{k}_{\parallel }}\cos k_{z}c,  \label{tb2}
\end{equation}
$\mathbf{k}_{\parallel }=\left( k_{x},k_{y}\right) $ refers to the in plane
momentum and 
\begin{equation}
\varepsilon _{\mathbf{k}_{\parallel }}=t_{\parallel }\left( 2+\cos
k_{x}a+\cos k_{y}a\right)
\end{equation}
is the in-plane dispersion. For convenience we included the constant shift $
2t_{\parallel }$ in the definition of $\varepsilon _{\mathbf{k}_{\parallel
}} $ to ensure that $\varepsilon _{\mathbf{k}_{\parallel }}\geq 0$. The last
term in Eq.(\ref{tb2}) refers to the inter-plane coupling, where the form
factor 
\begin{equation}
\gamma _{\mathbf{k}_{\parallel }}=\cos \frac{k_{x}a}{2}\cos \frac{k_{y}a}{2}
\end{equation}
describes the $\mathbf{k}_{\parallel }$-dependence of this coupling in the
BCT lattice. This $\mathbf{k}_{\parallel }$-dependence is a crucial aspect
of our theory.

For $t_{\parallel }$, $t_{\perp }>0$ and $t_{\parallel }>t_{\perp }/2$, Bose
Einstein condensation takes place at $\mathbf{Q}=\left( \pi /a,\pi
/a,k_{z}\right) $. Since $\gamma _{\mathbf{k}_{\parallel }}$ vanishes for $
\mathbf{k}_{\parallel }=\left( \pi /a,\pi /a\right) $, $E_{\mathbf{Q}}$ is
independent of $k_{z}$. The minimum of the dispersion is infinitely
degenerate as the $z$-component of the wave vector can take any value when
the $x$ and $y$ components are equal to $\pi /a$. In case of the ideal Bose
gas ($u=0$) this implies for $T=0$ that different layers decouple
completely. Only excitations at finite $T$ with in-plane momentum away from
the condensation point can propagate in the $z$-direction. This behavior
changes as soon as boson-boson interactions ($u>0$) are included. States in
the Bose condensate scatter and create virtual excitations above the
condensate that are allowed to propagate in the $z$-direction. These
excitations couple to condensate states in other layers\cite{Maltseva05}.
The condensed state of interacting bosons is then truly three dimensional,
even at $T=0$. This order by disorder argument for dimensional restoration
due to interactions does not apply in case of chemical potential tuned BEC.
In this case, the number of bosons at $T=0$ is strictly zero for $\mu <0$,
i.e. before BEC sets in. The absence of particles makes their interaction
mute and one can approach the QCP arbitrarily closely without coherently
coupling different layers.

From now on, we will measure the momentum relative to the wave vector $
\mathbf{Q}_{0}=(\pi /a,\pi /a,0)$: $\mathbf{q}=\mathbf{k}-\mathbf{Q}_{0}$,
such that BEC corresponds to a macroscopic occupation of a state with
vanishing in-plane momentum, $\mathbf{q}_{\parallel }=0$. Since we will
treat the inter-layer hopping, $t_{\perp }$, pertubativly, it is convenient
to rewrite $H_{B}$ using real space variables for the direction
perpendicular to the planes. 
\begin{eqnarray}
H_{B} &=&\sum_{\mathbf{q}_{\parallel },i}(\varepsilon _{\mathbf{q}
_{\parallel }}-\mu )a_{\mathbf{q}_{\parallel }i}^{\dagger }a_{\mathbf{q}
_{\parallel }i}^{\;}+u\sum_{\mathbf{x}_{\parallel },i}n_{\mathbf{x}
_{\parallel }i}n_{\mathbf{x}_{\parallel }i}  \notag \\
&+&t_{\perp }\!\!\!\sum_{\mathbf{q}_{\parallel },ij}\gamma _{\mathbf{q}
_{\parallel }}\eta _{ij}(a_{\mathbf{q}_{\parallel }i}^{\dagger }a_{\mathbf{q}
_{\parallel }j}^{\;}+h.c.),
\end{eqnarray}
The indices $i$,$j$ denote the different layers, with $\eta _{i,j}=1$ for
nearest neighbor layers while $\eta _{i,j}=0$ otherwise. Due to the shift of
momentum it holds 
\begin{equation}
\gamma _{\mathbf{q}_{\parallel }}=\sin \frac{q_{x}}{2}\sin \frac{q_{y}}{2}.
\end{equation}

We note that $H_{B}$ has a discrete $Z_{2}$--symmetry \cite
{Maltseva05,Rosch,Rosch2} : 
\begin{eqnarray}
q_{x} &\rightarrow &-q_{x},  \notag \\
a_{\mathbf{x}_{\parallel },i}^{\dagger } &\rightarrow &\left( -1\right)
^{i}a_{\mathbf{x}_{\parallel },i}^{\dagger }.  \label{Z2}
\end{eqnarray}
for all $i$. This is a local $Z_{2}$--symmetry with respect to the layer
index. In momentum space \ the last equation corresponds to $
q_{z}\rightarrow q_{z}+\pi /c$. The in-plane dispersion and the local
interaction trivially obey this symmetry. However, the inter-layer hopping
is only invariant with respect to this transformation since $\gamma _{
\mathbf{q}_{\parallel }}$ is odd \ with respect to either $q_{x}$ or $q_{y}$
. This discrete symmetry is therefore closely connected to the degeneracy of
the Bose condensate with respect to $q_{z}$. If we were to include an
additional inter-layer hopping term between neighboring planes with $\mathbf{
\ q}_{\parallel }$-independent hopping $t_{1}$, 
\begin{equation}
T_{1}=t_{1}\sum_{\mathbf{q}_{\parallel },ij}\eta _{ij}(a_{\mathbf{q}
_{\parallel }i}^{\dagger }a_{\mathbf{q}_{\parallel }j}^{\;}+h.c.),
\label{T1}
\end{equation}
we would break the $Z_{2}$-symmetry. In addition we would lift the
degeneracy of the Bose condensate to either $k_{z}=\pi /c$ or $k_{z}=0$,
depending on the sign of $t_{1}$. On the other hand, inclusion of a term 
\begin{equation}
T_{2}=t_{2}\sum_{\mathbf{q}_{\parallel },ij}\widetilde{\eta }_{ij}(a_{
\mathbf{q}_{\parallel }i}^{\dagger }a_{\mathbf{q}_{\parallel }j}^{\;}+h.c.),
\label{T2}
\end{equation}
that promotes boson hopping between second neighbors ($\widetilde{\eta }
_{i,i+2}=\widetilde{\eta }_{i+2,i}=1$ and $\widetilde{\eta }_{i,j}=0$
otherwise) would lift the degeneracy of the bose condensate, but without
breaking the $Z_{2}$-symmetry. We will see below that this leads to
important distinction between coherent coupling between nearest and next
nearest neighbor layers.

While the Bose condensed state for $\mu >0$ and the entire regime for $T>0$
is three dimensional, the decoupling for $\left( \mu <0,T=0\right) $ has
dramatic consequences. We show that the BEC transition temperature varies as 
\begin{equation}
T_{c}\propto \mu \ln \left( \frac{t_{\parallel }}{\mu }\right) /\ln \ln 
\frac{t_{\parallel }}{\mu }.  \label{Tc}
\end{equation}
$T_{c}$ $\propto \mu ^{2/d}$ holds instead for an isotropic Bose system in $
d>2$. Despite the fact that different layers are coupled at finite $T$ the
BEC-transition temperature, Eq.(\ref{Tc}), depends on $\mu $ just like the
Berezinskii-Kosterlitz-Thouless (BKT) transition temperature of a two
dimensional system \cite{FisherHohenberg88}.

The renormalization group (RG) calculation used to obtain this result (a
one-loop RG calculation in analogy to Ref.\cite{FisherHohenberg88,Sachdev94}
) shows that the finite temperature transition is a classical $3$-$d$ $XY$
transition, not a BKT transition. We conclude, therefore, that the $T=0$ QCP
of chemical potential tuned BEC with three dimensional dispersion, Eq.(\ref
{tb2}), is strictly two dimensional. The system then crosses over to be
three dimensional for $\mu >0$ or $T>0$, where the density of bosons becomes
finite and boson-boson interactions drive the crossover to $d=3$. The
transition temperature of this three dimensional BEC is given by the
two-dimensional result, Eq.(\ref{Tc}). It is important to stress that the
vanishing density for $\left( \mu <0,T=0\right) $ implies that these results
are not limited to weakly interacting bosons\cite{Sachdev}.

\section{Mean Field Theory}

\subsection{Phase Boundary}

Here we present a heuristic derivation of Eq.(\ref{Tc}) based on an approach
introduced by Popov \cite{Popov83} and further explored by Fisher and
Hohenberg \cite{FisherHohenberg88}: infrared divergencies are cut-off for
momenta $q<q_{0}\simeq \sqrt{\mu /t_{\parallel }}$, where $\mu $ is the
chemical potential. The main results of this approach were already given in
Ref.\cite{Batista07}. We will show how an effective coupling along the $z$
--axis appears when the interaction term of $H_{B}$ is taken into account.
For this purpose, we will approach the BEC-QCP from the disordered phase.
Since we are interested in the case of hard-core bosons, we will consider an
infinitely large on--site repulsive interaction $u\rightarrow \infty $.

\begin{figure}[tbh]
\vspace*{-0.0cm} \hspace*{0cm} 
\includegraphics[angle=0,width=9cm]{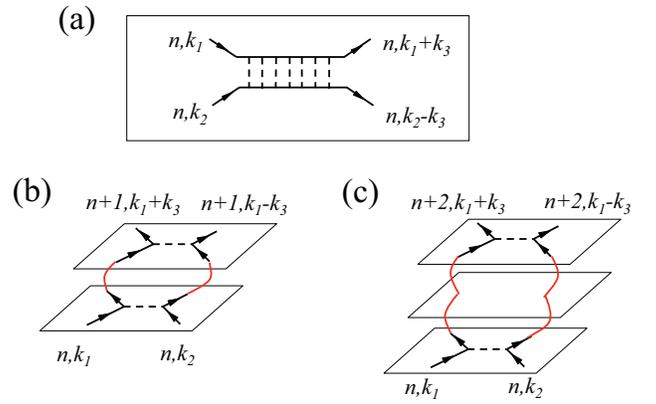}
\vspace{-2.0cm}
\caption{(a) Ladder diagrams that provide the dominant contribution to the
intra--layer scattering in the low density regime \protect\cite{Beliaev}.
(b) and (c) leading order diagrams that contribute to the coherent
inter--layer hoppings $t_{\perp ,1}^{\ast }$ and $t_{\perp ,2}^{\ast }$.}
\label{diagrams}
\end{figure}

While the interaction is local, scattering processes between bosons in
different layers generate effective non-local interactions at low energies
of the type 
\begin{equation}
{H_{\mathrm{int}}}=\frac{1}{2}\sum_{ijkl,\mathbf{x}_{\parallel }}v_{ijkl}( 
\mathbf{x}_{\parallel })a_{\mathbf{x}_{\parallel }i}^{\dagger }a_{{\mathbf{x}
}_{\parallel }j}^{\dagger }a_{\mathbf{x}_{\parallel }k}^{\;}a_{\mathbf{x}
_{\parallel }l}^{\;}.
\end{equation}
To leading order in the boson density $\rho $, the Fourier transform $v_{0}( 
\mathbf{q}_{\parallel })\equiv v_{iiii}(\mathbf{q}_{\parallel })$ of the
intra--layer effective interaction results from adding the ladder diagrams
shown in Fig.\ref{diagrams}a \cite{Beliaev}: 
\begin{equation}
\frac{1}{v_{0}(\mathbf{q}_{1\parallel }+\mathbf{q}_{2\parallel })}=\frac{1}{
4 }\int \frac{d^{2}q_{\parallel }}{4\pi ^{2}}\frac{1}{\varepsilon _{\mathbf{
\ q }_{\parallel }+{\mathbf{q}_{1\parallel }}}+\varepsilon _{{\mathbf{q}
_{2\parallel }-\mathbf{q}_{\parallel }}}},  \label{ladder}
\end{equation}
where 
\begin{equation}
\varepsilon _{{\mathbf{q}_{\parallel }}}=t_{\parallel }(2-\cos {q_{x}a}-\cos 
{q_{y}a})
\end{equation}
due to the shift of the in-plane momentum. The integral in Eq.(\ref{diagrams}
) diverges logarithmically in two dimensions for $\mathbf{q}_{1\parallel }, 
\mathbf{q}_{2\parallel }\rightarrow 0$. The effective interaction will be
logarithmically small in the low density limit. An heuristic way of deriving
a consistent mean field theory is to introduce the cut-off $q_{0}\sim \sqrt{
\mu /t_{\parallel }}$ \cite{Schick71,Popov83,FisherHohenberg88}: 
\begin{equation}
\frac{1}{v_{0}}=\frac{1}{4}\int_{q_{0}}^{\pi }\frac{d^{2}q_{\parallel }}{
4\pi ^{2}}\frac{1}{\epsilon _{\mathbf{q}_{\parallel }}}\propto \frac{\ln {\ 
\frac{t_{\parallel }}{\mu }}}{t_{\parallel }}.  \label{2dver}
\end{equation}

We proceed now to compute the inter-layer interactions \ $v_{ijkl}$ that are
generated by combining the intra--layer renormalized interaction, $v_{0}$,
with the inter-layer hopping term $t_{\perp }$. For this purpose, it is
convenient to expand the propagator in powers of the inter--layer hopping: 
\begin{eqnarray}
G_{ij}(q_{\parallel }) &=&g(q_{\parallel })\delta _{ij}+t_{ij}(q_{\parallel
})g^{2}(q_{\parallel })  \notag \\
&&+\sum_{l}t_{il}(q_{\parallel })t_{lj}(q_{\parallel })g^{3}(q_{\parallel
})+...,  \label{propexp}
\end{eqnarray}
where $t_{ij}(\mathbf{q}_{\parallel })=t_{\perp }\gamma _{\mathbf{q}
_{\parallel }}\eta _{i,j}$ and 
\begin{equation}
g(q_{\parallel })=\frac{1}{-i\omega _{n}+\frac{t_{\parallel }}{2}
q_{\parallel }^{2}-\mu }.  \label{gplane}
\end{equation}
is the intra--layer propagator for long wavelengths ($q_{\parallel }\ll 1$).
From now on, we \ measure in plane momenta in units of $2\pi /a$ where $a$
is the in-plane lattice constant (i.e. we set $a=1$) and \ work in the long
wavelength limit $q_{\parallel }\ll 1$. The leading order inter-layer
effective interactions that are relevant for inducing coherency along the $z$
-axis are $v_{iijj}\equiv v_{|i-j|}$ for $|i-j|=1$ (see Fig.\ref{diagrams}b)
and $|i-j|=2$ (see Fig.\ref{diagrams}c). Analyzing the corresponding ladder
diagrams yields: 
\begin{eqnarray}
v_{1}(p) &=&-v_{0}^{2}t_{\perp }^{2}\!\!\int \frac{d\omega d^{2}q}{\left(
2\pi \right) ^{3}}\gamma _{q}\gamma _{p-q}g^{2}(\ q)g^{2}(p-q),  \notag
\label{ints} \\
v_{2}(p) &=&-v_{0}^{2}t_{\perp }^{4}\!\!\int \frac{d\omega d^{2}q}{\left(
2\pi \right) ^{3}}\gamma _{q}^{2}\gamma _{p-q}^{2}g^{3}(q)g^{3}(p-q).  \notag
\end{eqnarray}
Performing the momentum and frequency integration with lower momentum cut
off $q_{0}$ and setting $p\rightarrow 0$ yields 
\begin{eqnarray}
v_{1} &=&-\frac{v_{0}^{2}t_{\perp }^{2}}{8\pi t_{\parallel }^{3}}\ln {\frac{
\pi }{q_{0}},}  \notag \\
v_{2} &=&-\frac{9v_{0}^{2}t_{\perp }^{4}}{128\pi t_{\parallel }^{5}}\ln {\ 
\frac{\pi }{q_{0}}.}  \label{v2}
\end{eqnarray}

The $v_{0}$, $v_{1}$ and $v_{2}$ processes generate the minimal number of
terms that have to be included in the low energy effective Hamiltonian in
order to provide a correct description of the critical properties of our
bosonic system in the low density limit. The expression for the new
interaction term in the low energy theory is 
\begin{equation}
{H_{\mathrm{int}}}=\frac{1}{2}\sum_{\mathbf{x}_{\parallel
}ij}\sum_{m=0}^{2}v_{m}\delta _{\left\vert i-j\right\vert ,m}\ a_{\mathbf{x}
_{\parallel }i}^{\dagger }a_{\mathbf{x}_{\parallel }i}^{\dagger }a_{\mathbf{
\ x }_{\parallel }j}a_{\mathbf{x}_{\parallel }j}.
\end{equation}
The \ $m=0$ term corresponds to intra--layer scattering vertex $v_{0}$. The
other terms with $m=1$, $2$ describe hopping of pairs of bosons from the
layer $i$ to the layer $j\pm m$. Now we perform the mean field decoupling
(we suppress the in-plane coordinate $\mathbf{x}_{\parallel }$ for
simplicity) 
\begin{eqnarray}
n_{i}n_{i} &\simeq &2\rho n_{i}-\rho ^{2},  \notag \\
a_{i}^{\dagger }a_{i}^{\dagger }a_{j}a_{j} &\simeq &a_{i}^{\dagger
}a_{j}^{\;}\langle a_{i}^{\dagger }a_{j}^{\;}\rangle +a_{i}^{\dagger
}a_{j}^{\;}\langle a_{i}^{\dagger }a_{j}^{\;}\rangle -\langle a_{i}^{\dagger
}a_{j}^{\;}\rangle ^{2},  \notag
\end{eqnarray}
where $\rho =\langle n_{\mathbf{x}_{\parallel },i}\rangle $. With this mean
field approximation we obtain an effective single particle Hamiltonian \
with dispersion 
\begin{equation}
E_{\mathbf{q}}^{\ast }=E_{\mathbf{q}}+2v_{1}\kappa _{1}\cos {q_{z}}
+2v_{2}\kappa _{2}\cos {2q_{z},}  \label{disp mf}
\end{equation}
with 
\begin{equation}
\kappa _{j}=\int \frac{d^{2}q_{\parallel }}{4\pi ^{2}}\left\langle a_{ 
\mathbf{q}_{\parallel },i}^{\dagger }a_{\mathbf{q}_{\parallel
},i+j}^{\;}\right\rangle ,  \label{kj2}
\end{equation}
and effective chemical potential 
\begin{equation}
\mu ^{\ast }=\mu -v_{0}\rho .
\end{equation}
The mean values $\langle a_{\mathbf{q}_{\parallel },i}^{\dagger }a_{\mathbf{
\ q }_{\parallel },i+j}^{\;}\rangle $ are given by: 
\begin{equation*}
\left\langle a_{\mathbf{q}_{\parallel },i}^{\dagger }a_{\mathbf{q}
_{\parallel },i+j}^{\;}\right\rangle =\int_{-\pi }^{\pi }\frac{dq_{z}}{2\pi }
\frac{\cos {(jq_{z})}}{e^{\beta \left( E_{\mathbf{q}}^{\ast }-\mu ^{\ast
}\right) }-1}.
\end{equation*}
It follows $\langle a_{\mathbf{q}_{\parallel },i}^{\dagger }a_{\mathbf{q}
_{\parallel },i+1}^{\;}\rangle =0$, a result that is a consequence of the
local $Z_{2}$ symmetry of $H$. This means that $\kappa _{1}$ may only become
non--zero when the $U(1)$ symmetry gets broken at the BEC transition. In
contrast, $\langle a_{\mathbf{x}_{\parallel },i}^{\dagger }a_{\mathbf{x}
_{\parallel },i+2}^{\;}\rangle $ is invariant under the discrete $Z_{2}$
-symmetry of $H$. Therefore this mean value is finite as long as the
concentration of bosons is finite. Although the term $2t_{\perp }\gamma _{ 
\mathbf{k}_{\parallel }}\cos {q_{z}}$ cancels at $\mathbf{q}_{\parallel }=0$
, it is crucial to keep it in order to obtain a finite value for $\kappa
_{2} $. Without this term, we have: $E_{\mathbf{q}}^{\ast }=E_{\mathbf{q+ 
\frac{ \pi }{2}{\hat{q}}_{z}}}^{\ast }$, which would imply $\langle a_{ 
\mathbf{q} _{\parallel },n}^{\dagger }a_{\mathbf{q}_{\parallel
},n+2}^{\;}\rangle =0$. The system undergoes a Bose-Einstein condensation
when the effective chemical potential, $\mu ^{\ast }$, becomes equal to
zero: 
\begin{equation}
\mu =v_{0}\rho (T_{c}).  \label{Tc23}
\end{equation}
In order to calculate $\rho (T_{c})$, we need to solve the integral (\ref
{kj2}) for $\kappa _{2}$ at $T=T_{c}$. We will assume that $\varepsilon _{ 
\mathbf{q}_{\parallel }}\gg 2v_{2}\kappa _{2}\cos {2q_{z}}$ for any $
q_{\parallel }\geq q_{0}$, and evaluate the expectation value $\left\langle
a_{\mathbf{q}_{\parallel },n}^{\dagger }a_{\mathbf{q}_{\parallel
},n+2}^{\;}\right\rangle $ in the limit $\kappa _{2}=0$. Below we verify
that this assumption is justified for small $t_{\perp }/t_{\parallel }$. It
follows 
\begin{equation}
\kappa _{2}\simeq \frac{Tt_{\perp }^{2}\ln \frac{2T}{t_{\parallel }q_{0}^{2}}
}{4\pi t_{\parallel }^{3}},
\end{equation}
where the logarithmic term contains again the lower momentum cut off $q_{0}$
. Without this lower cut off the mean field theory could not be properly
defined. This result is consistent with the above assumption that $
2v_{2}\kappa _{2}$ is small compared to $\varepsilon _{\mathbf{q}_{\parallel
}}$ if $q_{\parallel }\geq q_{0}$, since \ $\varepsilon _{\mathbf{q}
_{0}}\simeq t_{\parallel }\rho /\ln {\mu /t_{\parallel }}$ while $
2v_{2}\kappa _{2}\simeq \left( t_{\perp }/t_{\parallel }\right) ^{6}$\ $
\varepsilon _{\mathbf{q}_{0}}$. The last result was obtained using Eq.(\ref
{v2}) for $v_{2}$.

With $\kappa _{2}\neq 0$ for finite $T$, the effective dispersion $E_{ 
\mathbf{q}}^{\ast }$ of Eq.(\ref{disp mf}) becomes three dimensional.
Coherent motion of bosons within the planes and between planes is allowed.
While thermally excited bosons are needed for this coherent hopping to
emerge, its origin are quantum fluctuations that cause the non-local
interlayer interaction $v_{2}$. The quantum critical point at $\left(
T=0,\mu =0\right) $ is however purely two dimensional. We have a finite
coherent inter-layer coupling at the BEC momentum only for finite $T$ or in
the bose condensed state. This implies that the bose condensate itself is
three dimensional and that the universality class of the finite $T$
transition is $3D-XY$. However, the amplitude of this coherent coupling is
very small and the system will be effectively two dimensional until it is
very close to the transition. The width of the regime with three dimensional
fluctuations shrinks to zero as $T_{c}$ vanishes. \ This implies that the
magnitude of $T_{c}$ obtained from Eq.(\ref{Tc23}) is practically the same
as the magnitude of the Kostelitz-Thouless temperature $T_{KT}$. The
thermodynamic phase transition is however always of second order. Since the
effective coupling $v_{2}$ induced by order from disorder is irrelevant for
the quantum critical point (the phase transition induced by changing the
chemical potential at $T=0$), it is relevant for the classical phase
transition at $T=T_{c}$. Therefore, the dependence of $T_{c}$ on $\mu $ and $
\rho $ is given by the $d=2$ expressions: 
\begin{eqnarray}
T_{c} &\propto &t_{\parallel }\frac{\rho }{\ln {\ln {\rho ^{-1}}}},  \notag
\\
\mu &\propto &\frac{T_{c}}{\ln {t_{\parallel }/T_{c}}}.  \label{Tcres}
\end{eqnarray}

In addition we have the usual two dimensional expressions for the density as
function of temperature or chemical potential: 
\begin{eqnarray}
\rho (T=0,\mu ) &\propto &\mu \ln {\frac{\mu }{t_{\parallel }},}  \notag \\
\rho (T,\mu =0) &\propto &\frac{T}{t_{\parallel }}\ln {\ln {t_{\parallel
}/T. }}
\end{eqnarray}
The appeal of this mean field theory is its physical transparency and
technical simplicity. The introduction of the chemical potential as lower
cut off is however rather \textit{ad hoc }and it is unclear whether it is
justified for the problem at hand. In order to avoid these ambiguities we
developed a renormalization group approach that confirms the results of this
section (see below).

\subsection{Bond Ordering}

We will discuss now the bond ordering that accompanies the BEC. The $Z_2$
--symmetry (\ref{Z2}) is spontaneously broken below $T_c$ because according
to Eq.(\ref{kj2}) 
\begin{equation}
\kappa _{1} \simeq \frac{1}{4\pi ^{2}} \langle a_{ \mathbf{0} ,i}^{\dagger
}\rangle \langle a_{\mathbf{0},i+1}^{\;}\rangle,  \label{kappa1}
\end{equation}
becomes finite for a nonzero BEC order parameter $\langle a_{ \mathbf{0}
,i}^{\dagger }\rangle$. Moreover, $|\kappa_1|$ is maximized when the
relative phase between $\langle a_{ \mathbf{0},i}^{\dagger }\rangle=A
e^{\phi_i}$ and $\langle a_{\mathbf{0},i+1}^{\;}\rangle = A e^{\phi_{i+1}}$
is 0 or $\pi$ meaning that the inter--layer coupling favors any of these two
relative orientations below $T_c$: $E_{i,i+1} \propto \cos^2 {\
(\phi_{i+1}-\phi_{i})}$ \cite{Maltseva05}. In real space, this means that
the phase of a given site $\mathbf{x}_{\parallel}$ of the layer $i$, $\phi_{ 
\mathbf{x}_{\parallel},i}$, is parallel to the phase of two of its
nearest-neighbors on layer $i+1$ and antiparallel to the phase of the other
two. Consequently, the four bonds connecting a given site with its
nearest-neighbors on an adjacent layer become inequivalent below $T_c$,
i.e., there is a finite bond order parameter.

In principle, the bond ordering could appear at a critical temperature
higher than $T_{c}$. In that case there would be two thermodynamic phase
transitions instead of one. We will show now that there is only one phase
transition, i.e., that the bond order parameter becomes continuously nonzero 
\textit{only} below $T_{c}$. For this purpose, we introduce: 
\begin{equation}
\mu ^{\ast }=-2t_{\perp }^{\ast }-\delta {\mu }
\end{equation}
According to Eq.(\ref{disp mf}), the transition to the Bose condensed state
occurs for $\delta {\mu }=0$. Therefore, $\delta {\mu }$ measures the
deviation of the chemical potential from its critical value. By computing
the integral (\ref{kj2}) for $j=1$ we obtain: 
\begin{equation}
\kappa _{1}=\frac{T}{t_{\parallel }(2\pi )^{2}}\int_{0}^{2}\frac{(1-y)dy}{ 
\sqrt{y(2-y)}}\ln {\left[ 1-e^{-(\beta 2t_{\perp }^{\ast }y+\beta \delta \mu
+\beta \mu /t_{\parallel })}\right] }  \label{kj4}
\end{equation}
where $t_{\perp }^{\ast }=-v_{1}\kappa _{1}$ and we have used the heuristic
cut--off $q_{0}$. We do not expect any transition for $\delta \mu /\mu \gg 1$
because the temperature is much smaller than the excitation gap and the
number of bosons becomes exponentially small. Therefore, we will assume $
\delta \mu /\mu \ll 1$ that corresponds to the quantum critical regime with
the temperature (or the chemical potential) approaching the BEC--point from
the disordered side. If $t_{\perp }^{\ast }/\mu \ll 1$, we obtain: 
\begin{equation}
\kappa _{1}=\frac{T}{t_{\parallel }(2\pi )^{2}}\int_{0}^{2}\frac{(1-y)dy}{ 
\sqrt{y(2-y)}}\ln {(\beta 2t_{\perp }^{\ast }y+\beta \delta \mu +\beta \mu
/t_{\parallel })},
\end{equation}
that reduces to 
\begin{equation}
\frac{\delta \mu }{\mu }=1-\frac{Tv_{1}}{4\pi \mu t_{\parallel }}
\label{viol}
\end{equation}
after expanding the logarithm. Eq.(\ref{viol}) violates the original
assumption $\delta \mu /\mu \ll 1$ meaning that there is no solution of Eq.( 
\ref{kj4}) for any finite $\delta \mu $. This implies that the bond ordering
appears only below $T_{c}$.

\section{Renormalization Group Approach}

The mean field approach presented in the previous section was supplemented
by the introduction of a lower cut-off of otherwise infrared divergent \
terms in the perturbation theory. These divergencies result from the fact
that the two dimensional dilute Bose system is a quantum system at the upper
critical dimension. The natural approach to control these divergencies is a
renormalization group analysis.

In our renormalization group analysis of the model Eq.(\ref{Hb}) we start
from the action: 
\begin{eqnarray}
S_{\mathrm{bare}} &=&\sum_{ij}\int_{q}a_{q,i}^{\dagger }G_{ij}^{-1}\left(
q\right) a_{q,j} \\
&&+\frac{u}{2}\sum_{i}\int_{q_{1,2,3}}a_{q_{1},i}^{\dagger
}a_{q_{2},i}^{\dagger }a_{q_{3},i}a_{q_{1}+q_{2}-q_{3},i},  \notag
\end{eqnarray}
where 
\begin{equation}
G_{0ij}^{-1}\left( q\right) =g^{-1}(q)\delta _{ij}-t_{\perp }q_{x}q_{y}\eta
_{i,j}.
\end{equation}
with $g(q)$ of Eq.(\ref{gplane}). Here $q=\left( \mathbf{q}_{\parallel
},\omega _{n}\right) $ refers to the planar momentum $\mathbf{q}_{\parallel
}=\left( q_{x},q_{y}\right) $ and the bosonic Matsubara frequency $\omega
_{n}=2n\pi T$.\ We use the notation 
\begin{equation}
\int_{q}\ldots =T\int_{\left\vert \mathbf{q}_{\parallel }\right\vert
<\Lambda }\frac{d^{2}q_{\parallel }}{\left( 2\pi \right) ^{2}}\sum_{n}\ldots
\,.
\end{equation}
The upper cut off $\Lambda $ is determined by a length scale larger than the
inter--atomic spacing but \ much smaller than the correlation length. Thus $
\Lambda \simeq 1$ with our choice that the in-plane lattice constant $a=1$.
\ The upper momentum cut off $\Lambda $ yields an upper energy cut off of
order $t_{\parallel }$.

Although the inter-layer hopping $t_{\perp }$ is a marginal perturbation, it
is responsible for the emergence of new, non-local interactions, where
excited states of one layer propagate into another layer and couple to its
low energy states as it was already demonstrated in the mean field theory of
the previous section. We need to include such non-local couplings into the
effective action of the renormalization group analysis. Such non-local
interactions might cause, in turn, coherent motion of bosons even for $
\mathbf{q}_{\parallel }=0$. Thus, we need to further supplement the action
and allow for a coherent motion of bosons. This leads to the effective
action: 
\begin{eqnarray}
S &=&\sum_{ij}\int_{q}a_{q,i}^{\dagger }G_{ij}^{-1}\left( q\right) a_{q,j} \\
&&+\frac{1}{2}\sum_{ijkl}\int_{q_{1}q_{2}q_{3}}^{\Lambda
}v_{ijkl}a_{q_{1},i}^{\dagger }a_{q_{2},j}^{\dagger
}a_{q_{3},k}a_{q_{1}+q_{2}-q_{3},l}.  \notag
\end{eqnarray}
where 
\begin{equation}
G_{ij}^{-1}\left( q\right) =g^{-1}(q)\delta _{ij}-t_{ij}\left( \mathbf{q}
_{\parallel }\right) ,  \label{prop}
\end{equation}
with inter-layer hopping 
\begin{equation*}
t_{ij}\left( \mathbf{q}_{\parallel }\right) =\eta _{ij}\left( t_{1}+t_{\perp
}q_{x}q_{y}\right) +\widetilde{\eta }_{ij}t_{2}.
\end{equation*}
Thus, we include terms like $T_{1,2}$ in Eqs.\ref{T1} and \ref{T2}. In
particular, the hopping $t_{2}$ between next nearest neighbors is included
as it is the leading inter-layer boson hopping that does not violate the
above mentioned $Z_{2}$-symmetry. \ The hopping $t_{1}$ between neighboring
layers is included to explicitly demonstrate that it will not contribute to
coherent interlayer motion. At the beginning of the renormalization group
flow \ $S=S_{\mathrm{bare}}$ and it holds that 
\begin{equation}
t_{1}\left( l=0\right) =t_{2}\left( l=0\right) =0,
\end{equation}
and 
\begin{equation}
v_{ijkl}\left( l=0\right) =u\delta _{ij}\delta _{ik}\delta _{il}.
\end{equation}

\begin{figure}[tbh]
\vspace*{-0.5cm} \hspace*{-0.8cm} 
\includegraphics[angle=0,width=10cm]{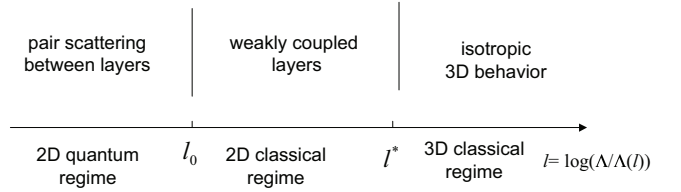} 
\vspace{-5.0cm}
\caption{Regimes of the renormalization flow.}
\label{flow}
\end{figure}

Before we derive one loop RG equations we discuss the various distinct
physical regimes indicated in Fig.\ref{flow}. The renormalization group
approach is controlled by the flow variable $l=\log \left( \Lambda / \Lambda
\left( l\right) \right)$. As usual in the regime of small but finite
temperature, $T$ is a relevant perturbation of the $T=0$ QCP and a crossover
to classical critical behavior occurs for $l>l_{0}$ when the renormalized
temperature becomes comparable with the upper energy cut off $t_{\parallel }$
: 
\begin{equation}
T\left( l_{0}\right) =t_{\parallel }.
\end{equation}
Excitations in the system with momentum larger than $\Lambda e^{-l_{0}}$
behave just like $T=0$ quantum excitations, while those with momentum below $
\Lambda e^{-l_{0}}$ are classical. As $T\rightarrow 0$, the crossover
variable $l_{0}\rightarrow \infty $ and, as expected, all degrees of freedom
are in the quantum regime. 

In addition to this quantum to classical crossover, the system undergoes a
dimensional crossover at a scale $l^{\ast }$ defined via 
\begin{equation}
t_{1,2}\left( l^{\ast }\right) =t_{\parallel },
\end{equation}
depending whether $t_{1}\left( l^{\ast }\right) $ or $t_{2}\left( l^{\ast
}\right) $ first reaches $t_{\parallel }$. In analogy to the quantum to
classical crossover, it holds that excitations with momentum larger than $
\Lambda e^{-l^{\ast }}$ behave quasi two dimensional while those with
momentum below $\Lambda e^{-l^{\ast }}$ are sensitive to a coherent
inter-layer coupling. We show that $l_{0}<l^{\ast }$ if the system is close
to the critical temperature, i.e. the dimensional crossover is driven by the
existence of thermal excitations in the system. However, quantum
fluctuations are nevertheless crucial for the dimensional crossover, as they
lead to non-local interactions $v_{ijkl}$ that are responsible for the
dimensional crossover once thermally excited bosons exist.

As long as the system is in the regime $l<l^{\ast }$, the renormalized
coherent inter-layer coupling is small. This has important implications for
the distinction between low and high energy degrees of freedom. For $
l<l^{\ast }$ we have to integrate out states with $\Lambda e^{-l}<\left\vert 
\mathbf{q}_{\parallel }\right\vert <\Lambda $, regardless of the momentum
perpendicular to $\mathbf{q}_{\parallel }$. Only once the RG flow enters a
three dimensional regime for $l>l^{\ast }$ is it sensible to distinguish low
and high energy modes with momentum $q_{z}$ perpendicular to the planes.
Then we integrate out states with $\Lambda e^{-l}<\sqrt{\mathbf{q}
_{\parallel }^{2}+q_{z}^{2}}<\Lambda $.

We first give the one loop renormalization group equations for $l<l^{\ast }$
. It holds 
\begin{eqnarray}
\frac{d \mu}{dl} &=& 2\mu -2\sum_{lm}v_{ilmi}\int_{k}^{>}G_{lm}\left( q
\right),  \notag \\
\frac{dt_{\perp }}{dl} &=& 0,  \notag \\
\frac{dt_{j}}{dl} &=&2t_{j}+2\sum_{lm}v_{i,l,m,i+j} \int_{k}^{>}G_{lm}\left(
q\right) ,  \notag \\
\frac{dT}{dl} &=&2T,  \label{RGsglp}
\end{eqnarray}
as well as 
\begin{eqnarray}
\frac{dv_{ijlm}}{dl} &=&-\sum_{stuv}v_{ijuv}v_{stlm}
\int_{q}^{>}G_{su}\left( q\right) G_{tv}\left( -q\right)  \notag \\
&&-4\sum_{stuv}v_{islu}v_{jtmv}\int_{q}^{>}G_{sv}\left( q\right)
G_{tu}\left( q\right) ,
\end{eqnarray}
where we use the short hand notation: 
\begin{equation}
\int_{q}^{>}...=\lim_{l\rightarrow 0}l^{-1}T\sum_{n}\int_{\Lambda
e^{-l}<\left\vert \mathbf{q}_{\parallel }\right\vert <\Lambda
}d^{2}q_{\parallel }...\,.
\end{equation}

For $l>l^{\ast }$ the renormalized inter-layer hopping element is comparable
to the in-plane hopping $t_{\parallel }$ and we are finally allowed to
perform a continuum's theory for the direction perpendicular to the layers
as well. Then, the problem is identical to the one of an isotropic three
dimensional Bose system 
\begin{eqnarray}
S &=&\int_{Q}^{\Lambda }\left( -i\omega _{n}+\frac{t_{\parallel }}{2}\left(
q_{\parallel }^{2}+q_{z}^{2}\right) -\mu \left( l^{\ast }\right) \right)
a_{Q}^{\dagger }a_{Q}  \notag \\
&&+\frac{1}{2}\int_{Q_{1}...Q_{4}}^{\Lambda }v_{iso}\left( l^{\ast }\right)
a_{Q_{1}}^{\dagger }a_{Q_{2}}^{\dagger }a_{Q_{3}}a_{Q_{1}+Q_{2}-Q_{3}}\ ,
\end{eqnarray}
where $Q=(\mathbf{q}_{\parallel },q_{z},\omega _{n})$ is a $(3+1)$
-dimensional vector that includes the momentum $q_{z}$. The initial values
for this flow are determined by the final values for the flow for $l<l^{\ast
}$. The isotropic boson interaction 
\begin{equation}
v_{iso}\left( l^{\ast }\right) =c\sum_{ijkl}v_{ijkl}\left( l^{\ast }\right) ,
\end{equation}
corresponds to the $q_{z}=0$ value of the coupling constants $v_{ijkl}$. For
short ranged couplings $v_{iso}$ it is dominated by the largest coupling
constant. The additional prefactor $c$, with lattice constant in the $z$
-direction, ensures that $v_{iso}$ is a three dimensional coupling constant
with dimension $\left[ \mathrm{length}\right] ^{d}\left[ \mathrm{energy} 
\right] $ for $d=3$.

\subsection{Quantum-Classical Crossover}

We first analyze the quantum to classical crossover at $l=l_{0}$. It is
useful to consider separately the behavior in the quantum regime, where the
renormalized temperature is small compared to the upper energy cut off $
T\left( l\right) <t_{\parallel }$ and the classical regime, where $T\left(
l\right) >t_{\parallel }$. We treat both regimes separately and \ assume $
T\left( l\right) \ll t_{\parallel }$ in the former and $T\left( l\right) \gg
t_{\parallel }$ in the latter regime and connect the flow of the various
coupling constants smoothly at $l_{0}$. This is essentially the approach
taken in Ref.\cite{FisherHohenberg88}.\ The analysis of Ref.\cite{Millis93}
demonstrates that the approach used here is fully consistent with results
obtained using a more careful analysis of the crossover behavior.

At $T=0$, the flow of the chemical potential and of the coherent inter-layer
boson hopping $t_{ij}$ are unaffected by the interaction between bosons.
This result is specific for the problem of dilute bosons with $\mu <0$ since
for $T=0$ 
\begin{equation}
\int_{q}^{>}G_{lm}\left( q\right) =0,
\end{equation}
as a result of the integration over frequency. Physically this is due to the
fact that the boson number vanishes for $T=0$ and $\mu <0$. This yields the
renormalization group equations: 
\begin{eqnarray}
\frac{d\mu (l)}{dl} &=&2\mu (l),  \notag \\
\frac{dt_{1,2}\left( l\right) }{dl} &=&2t_{1,2}\left( l\right) .
\label{flowq01}
\end{eqnarray}
At $T=0$, interaction do not affect $\mu $, $t_{1}$ and $t_{2}$. This only
happens once the system reaches the classical regime $l>l_{0}$. Since $
t_{1,2}\left( 0\right) =0$, it follows that $t_{1,2}\left( l\right) =0$ as
long as the system is in the quantum regime. Quantum fluctuations do not
induce a coherent hopping between layers. The inter-layer hopping $
t_{ij}\left( \mathbf{k}_{\parallel }\right) =\eta _{ij}t_{\perp }k_{x}k_{y}$
remains unchanged. Note, that the amplitude $t_{\perp }$ of this inter-layer
coupling is unchanged under renormalization. The flow of the chemical
potential is 
\begin{equation}
\mu \left( l\right) =\mu e^{2l}.
\end{equation}

Next we analyze the behavior of the interactions. \ At $T=0$ holds 
\begin{equation}
\int_{q}^{>}G_{sv}\left( q\right) G_{tu}\left( q\right) =0,
\end{equation}
which vanishes again because of the vanishing Boson density. We are left
with the analysis of 
\begin{equation}
\frac{dv_{ijlm}}{dl}=-\sum_{stuv}v_{ijuv}v_{stlm}\int_{q}^{>}G_{su}\left(
q\right) G_{tv}\left( -q\right) .
\end{equation}
In Appendix A we analyze this flow equation in the limit where the bare
inter-layer hopping $t_{\perp }$ is much smaller than the in-plane hopping $
t_{\parallel }$. Up to order $\left( t_{\perp }/t_{\parallel }\right) ^{4}$
we have to analyze the coupling for bosons in the same layer $v_{0}=v_{iiii}$
, in neighboring layers $v_{1}=v_{iijj}$ with $j=i\pm 1$ and second neighbor
layers $v_{1}=v_{iijj}$ with $j=i\pm 2$. \ We obtain at large $l$ ($l\gg
2\pi t_{\parallel }/u$): 
\begin{eqnarray}
v_{0}\left( l\right) &\simeq &\frac{2\pi t_{\parallel }a^{2}}{l}\left( 1- 
\frac{1}{2}\left( t_{\perp }/t_{\parallel }\right) ^{2}+\frac{3}{32}\left(
t_{\perp }/t_{\parallel }\right) ^{4}\right) ,  \notag \\
v_{1}\left( l\right) &\simeq &-\frac{\pi t_{\parallel }a^{2}}{2l}\left(
\left( t_{\perp }/t_{\parallel }\right) ^{2}-\left( t_{\perp }/t_{\parallel
}\right) ^{4}\right) ,  \notag \\
v_{2}\left( l\right) &\simeq &-\frac{\pi t_{\parallel }a^{2}}{l}\ \frac{5}{
32}\left( t_{\perp }/t_{\parallel }\right) ^{4}.  \label{finq}
\end{eqnarray}
It is important to keep in mind that these results were obtained with the
assumption that initially $v_{0}\left( l=0\right) =u$ is the \ only coupling
constant. The inter-layer interactions $v_{1}$ and $v_{2}$ result from
multiple scatterings in distinct layers where virtual bosons propagate
between layers. Thus, we find that there is no coherent coupling between
layers in the quantum regime i.e $t_{1,2}\left( 0\right) =0$. On the other
hand, we do find that non-local interactions, that couple different layers,
emerge. This is fully consistent with the finding of Refs.\cite
{Maltsevera05,Yildirim96}.

At finite $T$, the flow in the quantum regime stops at 
\begin{equation}
l_{0}=\frac{1}{2}\log \left( t_{\parallel }/T\right) .
\end{equation}
For $l>l_{0}$ thermal, as opposed to quantum fluctuations, come into play.
The initial values for the subsequent flow are of course the final values of
the RG flow of the quantum regime: $\mu \left( l_{0}\right) =\mu
e^{2l_{0}}=t_{\parallel }\frac{\mu }{T}$ and $v_{i}\left( l_{0}\right) $
where the $v_{i}\left( l\right) $ are given in Eq.(\ref{finq}).

\subsection{Dimensional Crossover}

The RG flows for $l>l_{0}$ continues to be two dimensional, as no coherent
inter-layer was generated in the quantum regime. As discussed above we will
now analyze the flow equations as if the problem was purely classical, i.e.
we include solely the lowest Matsubara frequency in the evaluation of the
Feynman diagrams. In this case temperature only enters the flow equations in
the combination 
\begin{equation}
w_{i}\left( l\right) =\frac{T\left( l\right) }{t_{\parallel }}v_{i}\left(
l\right) .
\end{equation}
Thus, it is convenient to use $w_{i}\left( l\right) $ in what follows. The
leading order flow equations of $w_{i}\left( l\right) $ are 
\begin{equation}
\frac{dw_{i}\left( l\right) }{dl}=2w_{i}\left( l\right) ,  \label{flowclassu}
\end{equation}
with solution 
\begin{equation}
w_{i}\left( l\right) =w_{i}\left( l_{0}\right) e^{2\left( l-l_{0}\right)
}=v_{i}\left( l_{0}\right) e^{2\left( l-l_{0}\right) }.  \label{wl}
\end{equation}
The coupling constants $w_{i}\left( l\right) $ are relevant. This is a
consequence of the fact that the upper critical dimension of the classical
regime is $d_{u,class.}=4$ as opposed to $d_{u,qu.}=2$ for the zero
temperature quantum regime. If we wanted to determine the critical exponents
of the classical phase transition, we would have to include higher order
terms. As pointed out by Millis\cite{Millis93}, it is not necessary to
include these higher order terms if one only wants to determine the value of
the transition temperature: At low $T$, the coupling constants $v_{i}\left(
l\right) $ decrease for large $l_{0}$ as follows from Eq.(\ref{finq}). Thus,
the initial values $w_{i}\left( l_{0}\right) $ of the classical flow are
small. While the interactions become relevant for $l>l_{0}$ corrections to
Eq.(\ref{flowclassu}) remain negligible unless the flow enters the actual
critical regime. However, in our case the flow only enters the critical
regime after the dimensional crossover. Thus, we can, for the moment, safely
neglect corrections beyond Eq.(\ref{flowclassu}).

As shown in Appendix B, the RG flow equations for the coherent hopping
elements and the chemical potential in the classical regime are 
\begin{eqnarray}
\frac{d\mu }{dl} &=&2\mu -\frac{2}{\pi }w_{0},  \notag \\
\frac{dt_{1}}{dl} &=&2t_{1}\ ,  \notag \\
\frac{dt_{2}}{dl} &=&2t_{2}+\frac{t_{\perp }^{2}}{\pi t_{\parallel }^{2}}\
w_{2}.  \label{flowhop}
\end{eqnarray}
It immediately follows that $t_{1}\left( l\right) =0$ since $t_{1}\left(
l_{0}\right) =0$. No coherent nearest neighbor hopping $t_{1}$ is being
generated by the mechanism we describe. This is a consequence of the
discussed $Z_{2}$-symmetry. A finite value for $t_{1}$ corresponds to a
broken $Z_{2}$ symmetry. \ However, the second neighbor coupling $t_{2}$
flows to a finite value even if its initial value vanishes. If we use $
w_{i}\left( l\right) $ of Eq.(\ref{wl}) with initial values $v_{i}\left(
l_{0}\right) $ from Eq.(\ref{finq}) it follows 
\begin{eqnarray}
\frac{d\mu }{dl} &=&2\mu -\frac{g_{0}t_{\parallel }\ }{l_{0}}\ e^{2\left(
l-l_{0}\right) },  \notag \\
\frac{dt_{2}}{dl} &=&2t_{2}-\frac{g_{2}t_{\parallel }}{l_{0}}\ e^{2\left(
l-l_{0}\right) },
\end{eqnarray}
where 
\begin{eqnarray}
g_{0} &=&4\left( 1-\frac{1}{2}\left( t_{\perp }/t_{\parallel }\right) ^{2}+ 
\frac{3}{32}\left( t_{\perp }/t_{\parallel }\right) ^{4}\right) ,  \notag \\
g_{2} &=&\frac{5}{32}\left( t_{\perp }/t_{\parallel }\right) ^{6}.
\end{eqnarray}
The solutions of these differential equations are 
\begin{eqnarray}
\mu \left( l\right) &=&e^{2\left( l-l_{0}\right) }\left( \mu \left(
l_{0}\right) -\frac{g_{0}t_{\parallel }\ }{l_{0}}\left( l-l_{0}\right)
\right) ,  \notag \\
t_{2}\left( l\right) &=&-e^{2\left( l-l_{0}\right) }\frac{g_{2}t_{\parallel
}\left( l-l_{0}\right) }{l_{0}}.
\end{eqnarray}
In the last equation we already took into account that the initial value of
the coherent hopping vanishes: $t_{2}\left( l_{0}\right) =0$. The
dimensional crossover takes place at $l^{\ast }$ where $\left\vert
t_{2}\left( l^{\ast }\right) \right\vert \simeq t_{\parallel }$, which
corresponds to 
\begin{equation}
e^{2\left( l^{\ast }-l_{0}\right) }\frac{g_{2}\left( l^{\ast }-l_{0}\right) 
}{l_{0}}=1.
\end{equation}
For large $l^{\ast }-l_{0}$ this is equivalent to 
\begin{equation}
e^{2\left( l^{\ast }-l_{0}\right) }\simeq \frac{l_{0}}{g_{2}\log l_{0}/g_{2}}
.
\end{equation}
This yields 
\begin{equation}
w_{0}\left( l^{\ast }\right) =v_{0}\left( l_{0}\right) e^{2\left( l^{\ast
}-l_{0}\right) }=v_{0}\left( l_{0}\right) \frac{l_{0}}{g_{2}\log l_{0}/g_{2}}
\end{equation}
as well as 
\begin{equation}
\mu \left( l^{\ast }\right) =\left( \frac{l_{0}}{g_{2}\log l_{0}/g_{2}}\mu
\left( l_{0}\right) -\frac{g_{0}\ }{g_{2}}t_{\parallel }\right)
\end{equation}
Inserting $l_{0}=\frac{1}{2}\log \left( \varepsilon _{0}/T\right) $ $\ $and $
\mu \left( l_{0}\right) $ gives 
\begin{equation}
w_{0}\left( l^{\ast }\right) =\frac{g_{0}\pi t_{\parallel }}{2}\frac{1}{
g_{2}\log \left( \frac{1}{2}\log \left( t_{\parallel }/T\right)
/g_{2}\right) }
\end{equation}
for the value of the coupling constant at the end of the two dimensional
flow and 
\begin{equation}
\mu \left( l^{\ast }\right) =t_{\parallel }\left( \frac{\frac{1}{2}\log
\left( t_{\parallel }/T\right) }{g_{2}\log \left( \frac{1}{2}\log \left(
t_{\parallel }/T\right) /g_{2}\right) }\frac{\mu }{T}-\frac{g_{0}\ }{g_{2}}
\right)
\end{equation}
for the corresponding chemical potential. As pointed out above, for $
l>l^{\ast }$, the RG probes energies sufficiently low to be sensitive to the
three dimensional character of the system. These final values of the
combined quantum and classical two dimensional flow become the initial value
of the three dimensional flow. Since always holds $l^{\ast }>l_{0}$, it
follows that this three dimensional flow is always in the classical regime.

\subsection{Flow in the Three-Dimensional Classical Regime}

The final regime of the flow is in the classical three dimensional regime.
The flow equations are the usual ones for an isotropic three dimensional
classical bosonic system, i.e. for an two component $\varphi ^{4}$ or $XY$
model. The condition for the critical temperature is that the initial values
for the flow of this three dimensional classical flow obey: 
\begin{equation}
\mu \left( l^{\ast }\right) \simeq \ w\left( l^{\ast }\right) .  \label{crit}
\end{equation}
This ensures that the flow is on the critical surface and the system is
close to the critical temperature. An alternative way to interpret this
condition was given in Ref.\cite{Millis93} where it was shown that Eq.(\ref
{crit}) is equivalent to the Ginzburg criterion for the onset of critical
fluctuations. Whenever a system is in the Ginzburg regime of classical
critical fluctuations, it is very close to the actual critical temperature.
The detailed analysis inside this regime is the usual one for a $d=3$
classical $XY$ model and does not need be reproduced here. We are more
interested in the value of the transition temperature at low $T$. We use our
previous results for the initial values $\mu \left( l^{\ast }\right) $ and $
w\left( l^{\ast }\right) $ of the three dimensional flow to analyze the
condition Eq.(\ref{crit}) \ and obtain 
\begin{eqnarray}
\mu &=&T_{c}\frac{2g_{0}\left( \frac{\pi }{2}+\log \left( \frac{1}{2g_{2}}
\log \left( t_{\parallel }/T_{c}\right) \right) \right) }{\log \left(
\varepsilon _{0}/T_{c}\right) }.  \notag \\
&\simeq &8T_{c}\frac{\log \left( \frac{16}{5\varepsilon ^{6}}\log \left(
t_{\parallel }/T_{c}\right) \right) }{\log \left( \varepsilon
_{0}/T_{c}\right) }.
\end{eqnarray}
Solving this result for $T_{c}$ with logarithmic accuracy yields the
transition temperature as function of chemical potential, as given in Eq.( 
\ref{Tc}). The phase diagram that results from our RG analysis is
represented in Fig.\ref{phase_diagram}.

\section{Summary}

In summary, we have shown that inter--layer frustration reduces the
effective dimensionality of a BEC quantum phase transition induced by a
change of the chemical potential. The BEC-QCP exhibits 2D quantum critical
fluctuations that dominate over an extended region of the phase diagram. The
phase boundary between the disordered and ordered phase extends to finite
temperatures although the universality class of the transition changes from
BEC in 2+2 dimensions at $T=0$ to 3D--XY at finite $T$. For $T>0$, there is
a crossover from the 2D quantum critical to a 2D classical regime as the
system approaches the phase boundary from disordered side. The dimensional
crossover occurs within the classical regime as the system gets even closer
to the phase boundary (see Figs.\ref{phase_diagram} and \ref{flow}).

The BEC ordering is accompanied by bond ordering that results from a
spontaneous breaking of the $Z_2$--symmetry discussed in section II. Both,
the BEC and the bond order parameters increase continuously from zero for $
\mu \to \mu_c$. A finite bond--order parameter induces a finite hopping
between nearest--neighbor layers that vanishes at the phase boundary
together with the bond ordering.

Although according to our results the thermodynamic phase transition always
belongs to the 3D--XY universality class, this transition becomes more
quasi--2D like as the system approaches the 2D BEC-QCP. Besides the
consequences that were already discussed in the paper, like the peculiar
behavior of $T_c(\mu)$ given by Eq.(\ref{Tc}), this observation has
implications for the $T$--dependence of any thermodynamic quantity for $T
\to T_c$ and $\mu \gtrsim \mu_c(T=0)$ given the dimensional crossover
predicted by our RG calculation.

The dimensional reduction at a QCP that we discussed in this paper can be
experimentally verified in real quantum magnets such as BaCuSi$_2$O$_6$ \cite
{Suchitra06,Batista07}. For quantum magnets, the chemical potential
corresponds to a magnetic field applied along the symmetry axis while the
particle density corresponds ot the magnetization per site. Therefore, the
quantum phase transition discussed in this paper corresponds to the
suppression of magnetic XY--ordering by the application of a magentic field
that saturates the moments along the Z--direction. Although we discussed the
case of a BCT lattice, our result can be applied to more general layered
structures with frustrated inter--layer coupling.

\section{Acknowledgement}

We thank A. J. Millis, N. Prokof'ev and M. Vojta for helpful discussions and
to M. Vojta for pointing out Ref.\cite{Yildirim96}. LANL is supported by US
DOE under Contract No. W-7405-ENG-36. Ames Laboratory, is supported by US
DOE under Contract No. W-7405-Eng-82.  

\appendix

\section{Appendix}

\subsection{Flow in the $2D$-quantum regime}

In this appendix we derive the result, Eq.(\ref{finq}) for the interactions $
v_{i}$ in the regime $l<l_{0}$ prior to the quantum to classical crossover,
by solving flow equations 
\begin{equation}
\frac{dv_{ijlm}}{dl}=-\sum_{stuv}v_{ijuv}v_{stlm}\int_{k}^{>}G_{su}\left(
k\right) G_{tv}\left( -k\right) .
\end{equation}
We derive these results by expanding with respect to the ratio $t_{\perp
}/t_{\parallel }$ \ of the hopping elements perpendicular and parallel to
the layers. Thus we expanded \ the propagator $G_{ij}\left( k\right) $ of
Eq.(\ref{prop}) in powers of the inter-layer hopping, see Eq.(\ref{propexp}
). In perturbation theory in $t_{\perp }$, it always holds that $i=j$, $l=m$
, $u=v$ , and $s=t$. Thus we obtain (to simplify the notation we use $
v_{ij}=v_{iijj}$): 
\begin{equation}
\frac{dv_{ij}}{dl}=-\sum_{st}v_{it}v_{sj}\int_{q}^{>}G_{st}\left( q\right)
G_{st}\left( -q\right) .
\end{equation}
Including terms up to order $t_{\perp }^{4}$ it follows: 
\begin{eqnarray}
\frac{dv_{ij}}{dl} &=&-\sum_{s}v_{is}v_{sj}\left( A^{\left( 0\right)
}+4B^{\left( 2\right) }\right) \ -\sum_{st}v_{it}v_{sj}\eta _{st}A^{\left(
2\right) }  \notag \\
&&-\sum_{st}v_{it}v_{sj}\sum_{l}\eta _{sl}\eta _{lt}A^{\left( 4\right) },
\end{eqnarray}
where 
\begin{eqnarray*}
A^{\left( n\right) } &=&\frac{t_{\perp }^{n}}{l}\int_{q}^{>}\gamma \left(
q\right) ^{n}g\left( q\right) ^{\frac{n+2}{2}}\ g\left( -q\right) ^{\frac{
n+2 }{2}}, \\
B^{\left( 2\right) } &=&\frac{t_{\perp }^{2}}{l}\int_{q}^{>}\gamma \left(
q\right) ^{2}g\left( q\right) \ g\left( -q\right) ^{3},
\end{eqnarray*}
with $\gamma \left( k\right) =k_{x}k_{y}$ and $g\left( q\right) $ of Eq.(\ref
{gplane}). Performing the frequency and momentum sums yields $A^{\left(
0\right) }=(2\pi t_{\parallel })^{-1}$, $A^{\left( 2\right) }=\frac{t_{\perp
}^{2}}{t_{\parallel }^{2}}\left( 8\pi t_{\parallel }\right) ^{-1}$, $
B^{\left( 2\right) }=A^{\left( 2\right) }/2$ as well as $A^{\left( 4\right)
}=t_{\perp }^{4}/t_{\parallel }^{4}\times 9\left( 128\pi t_{\parallel
}\right) ^{-1}$. Only terms with $j=i\pm 1$ and $j=i\pm 2$ are being
generated at fourth order in $t_{\perp }$. We will then introduce three
different coupling constants, $v_{0}=v_{ii}$, $v_{1}=v_{i,i\pm 1}$ and $
v_{2}=v_{i,i\pm 2}$. It will turn out to be crucial to include $v_{2}$ in
addition to the leading non-local coupling $v_{1}$. Performing the lattice
sums yields explicit flow equations for the three coupling constants. If we
now keep in mind that due to the initial conditions $v_{1}\left( l=0\right)
=v_{2}\left( l=0\right) =0$ vertices with $v_{1}$ are at least of order $
t_{\perp }^{2}$ and vertices with $v_{2}$ are at least of order $t_{\perp
}^{4}$ $\ $\ we can restrict the flow equations to \ fourth order in $
t_{\perp }$: 
\begin{eqnarray}
\frac{dv_{0}}{dl} &=&-v_{0}^{2}\widetilde{A}^{\left( 0\right) }\
-2v_{1}^{2}A^{\left( 0\right) }-4v_{0}v_{1}A^{\left( 2\right) }, \\
\frac{dv_{1}}{dl} &=&-2v_{0}v_{1}\left( A^{\left( 0\right) }+4B^{\left(
2\right) }\right) \ -v_{0}^{2}A^{\left( 2\right) },  \notag \\
\frac{dv_{2}}{dl} &=&-\left( 2v_{0}v_{2}+v_{1}^{2}\right) A^{\left( 0\right)
}\ -2v_{0}v_{1}A^{\left( 2\right) }-v_{0}^{2}A^{\left( 4\right) },  \notag
\end{eqnarray}
with $\widetilde{A}^{\left( 0\right) }=A^{\left( 0\right) }+4B^{\left(
2\right) }+2A^{\left( 4\right) }$. For large $l$ we expect a decay of the
coupling constants according to $v_{\alpha }\left( l\right) \ \propto l^{-1}$
. Thus we assume 
\begin{equation}
v_{\alpha }\left( l\right) =\frac{h_{\alpha }\left( l\right) }{l},
\label{vvsh}
\end{equation}
and analyze the flow equations for $h_{\alpha }\left( l\right) $. For large
enough $l$ we can determine the amplitudes of the coupling constants from $
\frac{dh_{\alpha }}{dl}=0$, leading to the algebraic equations 
\begin{eqnarray}
h_{0}\ &=&h_{0}^{2}\widetilde{A}^{\left( 0\right) }\ +2h_{1}^{2}A^{\left(
0\right) }+4h_{0}h_{1}A^{\left( 2\right) }, \\
h_{1} &=&2h_{0}h_{1}\left( A^{\left( 0\right) }+4B^{\left( 2\right) }\right)
\ +h_{0}^{2}A^{\left( 2\right) },  \notag \\
h_{2} &=&\left( 2h_{0}h_{2}+h_{1}^{2}\right) A^{\left( 0\right) }\
+2h_{0}h_{1}A^{\left( 2\right) }+h_{0}^{2}A^{\left( 4\right) }.  \notag
\end{eqnarray}
We can solve this system of equations once again by expanding with respect
to the small parameter 
\begin{equation}
\alpha =A^{\left( 2\right) }/A^{\left( 0\right) }=\frac{1}{4}\left( t_{\perp
}/t_{\parallel }\right) ^{2},
\end{equation}
keeping in mind that $A^{\left( 4\right) }/A^{\left( 0\right) }=\frac{9}{4}
\alpha ^{2}$. It follows 
\begin{eqnarray}
h_{0} &=&2\pi t_{\parallel }a^{2}\left( 1-2\alpha +\ \frac{7}{4}\alpha
^{2}\right) ,  \notag \\
h_{1} &=&-2\pi t_{\parallel }a^{2}\left( \alpha -4\alpha ^{2}\right) , 
\notag \\
h_{2} &=&-2\pi t_{\parallel }a^{2}\frac{5}{4}\alpha ^{2}.
\end{eqnarray}
Inserting these results into Eq.(\ref{vvsh}) yields the result Eq.(\ref{finq}
).

\subsection{Flow equations in the $2D$-classical regime\protect\bigskip}

As discussed in the main text, in the two dimensional classical regime we
concentrate of the flow equations of the chemical potential and coherent
hopping elements. We start from the general RG equations given in Eq.(\ref
{RGsglp}). Using the fact that $v_{ijkl}$ has only three nonvanishing
contributions $v_{m}=v_{iijj}$ \ with $j=i\pm m$ and $m=0$ (same layer), $
m=1 $, neighboring layers and $m=2$ (second neighbor layers). Inserting this
result into Eq.(\ref{RGsglp}) yields:

\begin{eqnarray}
\frac{d\mu }{dl} &=&2\mu -2v_{0}\int_{k}^{>}G_{ii}\left( k\right) ,  \notag
\\
\frac{dt_{1}}{dl} &=&2t_{1}+2v_{1}\int_{k}^{>}G_{ii+1}\left( k\right) , 
\notag \\
\frac{dt_{2}}{dl} &=&2t_{2}+2v_{2}\int_{k}^{>}G_{ii+2}\left( k\right) .
\label{flowint}
\end{eqnarray}
It holds up to second order in $t_{\perp }$: 
\begin{eqnarray}
G_{ii}\left( k\right) &\simeq &g\left( k\right) +2t_{\perp }^{2}\gamma
\left( k\right) ^{2}g\left( k\right) ^{3},  \notag \\
G_{ii+1}\left( k\right) &\simeq &t_{\perp }\gamma \left( k\right) g\left(
k\right) ^{2}\ ,  \notag \\
G_{ii+2}\left( k\right) &\simeq &\ t_{\perp }^{2}\gamma \left( k\right)
^{2}g\left( k\right) ^{3}.  \label{Gexp}
\end{eqnarray}
Here we ignored effects due to $t_{1}$ and $t_{2}$ as those will only be of
higher order in $t_{\perp }/t_{\parallel }$. This enables us to perform the
shell integration 
\begin{eqnarray}
\int_{k}^{>}G_{ii}\left( k\right) &=&\frac{T\left( l\right) }{\pi
a^{2}t_{\parallel }}\left( 1+\frac{t_{\perp }^{2}}{t_{\parallel }^{2}}
\right) ,\   \notag \\
\int_{k}^{>}G_{ii+1}\left( k\right) &=&0,  \notag \\
\int_{k}^{>}G_{ii+2}\left( k\right) &=&\frac{T\left( l\right) }{2\pi
a^{2}t_{\parallel }}\frac{t_{\perp }^{2}}{t_{\parallel }^{2}}\ ,
\end{eqnarray}
where we only included the zeroth's Matsubara frequency in the classical
regime. The contribution for the nearest neighbor coupling vanishes since $
\int_{k}\gamma \left( k\right) =0$, an effect caused by the $Z_{2}$-symmetry
of the Hamiltonian. \ Inserting these results into Eq.(\ref{flowint}) yields
Eq.(\ref{flowhop}). The solution of the flow equations then yields values
for the second neighbor hopping small by $\left( t_{\perp }/t_{\parallel
}\right) ^{6}$, justifying our assumption to neglect $t_{2}$ in the right
hand side of Eq.(\ref{Gexp}). It is also important to notice that including
terms with coherent neighbor hopping $t_{1}$ in Eq.(\ref{Gexp}) and self
consistently solving the RG equation for $t_{1}$ still yields $t_{1}=0$ on
the disordered side of the phase transition.

\end{document}